\documentclass[aps,prl,reprint,superscriptaddress]{revtex4-1}

\usepackage[dvips]{epsfig}

\usepackage[sort&compress]{natbib}

\begin{document}

\title{Two-photon interference using background-free quantum frequency conversion of single photons from a semiconductor quantum dot}

\author{Serkan Ates} \email{serkan.ates@nist.gov}
\affiliation{Center for Nanoscale Science and Technology,
National Institute of Standards and Technology, Gaithersburg, MD
20899, USA}
\affiliation{Maryland NanoCenter, University of Maryland, College
Park, MD}
\author{Imad Agha}
\affiliation{Center for Nanoscale Science and Technology,
National Institute of Standards and Technology, Gaithersburg, MD
20899, USA}
\affiliation{Maryland NanoCenter, University of Maryland, College
Park, MD}
\author{Angelo Gulinatti}
\affiliation{Politecnico di Milano, Dipartimento di Electronica e
Informazione, Piazza da Vinci 32, 20133 Milano, Italy}
\author{Ivan Rech}
\affiliation{Politecnico di Milano, Dipartimento di Electronica e
Informazione, Piazza da Vinci 32, 20133 Milano, Italy}
\author{Matthew T. Rakher}
\affiliation{Center for Nanoscale Science and Technology, National
Institute of Standards and Technology, Gaithersburg, MD 20899, USA}
\author{Antonio Badolato}
\affiliation{Department of Physics and Astronomy, University of
Rochester, Rochester, New York 14627, USA}
\author{Kartik Srinivasan}\email{kartik.srinivasan@nist.gov}
\affiliation{Center for Nanoscale Science and Technology, National
Institute of Standards and Technology, Gaithersburg, MD 20899, USA}
\date{\today}

\date{\today}

\begin{abstract}

We show that quantum frequency conversion (QFC) can overcome the
spectral distinguishability common to inhomogeneously broadened
solid-state quantum emitters. QFC is implemented by combining single
photons from an InAs quantum dot (QD) at 980\,nm with a 1550~nm pump
laser in a periodically-poled lithium niobate (PPLN) waveguide to
generate photons at 600\,nm with a signal-to-background ratio
exceeding 100:1. Photon correlation and two-photon interference
measurements confirm that both the single photon character and
wavepacket interference of individual QD states are preserved during
frequency conversion. Finally, we convert two spectrally separate QD
transitions to the same wavelength in a single PPLN waveguide and
show that the resulting field exhibits non-classical two-photon
interference.

\end{abstract}



\maketitle

Quantum frequency conversion (QFC)~\cite{ref:Kumar} is a potentially
crucial resource in interfacing photonic quantum systems operating
at disparate frequencies. Such a hybrid quantum system could, for
example, combine robust and stable quantum light sources based on
solid-state emitters~\cite{ref:Shields_NPhot} with broadband quantum
memories based on dense atomic ensembles~\cite{ref:Reim_NatPhot10}
to enable entanglement distribution in a long-distance quantum
network~\cite{ref:Sangouard_PRA}. QFC has been enabled by the
development of high-efficiency frequency conversion
techniques~\cite{ref:Fejer_IEEE,ref:Gnauck}, and been demonstrated
in experiments showing that the quantum character of a light field
was preserved during the
process~\cite{ref:Huang_Kumar_PRL,ref:Giorgi_PRL,ref:Tanzili_Zbinden,ref:Rakher_NPhot_2010,ref:McGuinnes_PRL10,ref:Ikuta_Imoto,ref:Zaske}.
It can be particularly valuable for solid-state quantum emitters, as
prominent systems like semiconductor quantum
dots~\cite{ref:Shields_NPhot} and nitrogen vacancy centers in
diamond~\cite{ref:Kurtsiefer} exhibit significant inhomogeneous
broadening. Thus, although these systems are in principle scalable,
applications which require identical quantum light sources need a
mechanism to bring spectrally disparate sources into
resonance~\cite{ref:Flagg_PRL10,ref:patel_Nphot2010,ref:Lettow,ref:Bernien_Hanson_PRL,ref:Sipahigil}.
Unlike previous demonstrations, in which techniques such as
strain/optical/electric fields were applied, QFC can fulfill this
role~\cite{ref:Takesue} without requiring direct modification of the
sources themselves.

Here, we demonstrate nearly background-free QFC, which we use to
enable experiments examining photon statistics and two-photon
interference of single photons from a semiconductor quantum dot.
Compared to previous telecommunications (1300\,nm) to visible
(710\,nm) conversion~\cite{ref:Rakher_NPhot_2010}, we work with
quantum dots (QDs) emitting in the well-studied 900\,nm to 1000\,nm
wavelength range~\cite{ref:Shields_NPhot}, and convert their single
photon emission to 600 nm, a wavelength region in which Si single
photon avalanche diodes (SPADs) offer a combination of quantum
efficiency and timing resolution that is currently unavailable in
the 980~nm band~\cite{ref:Ghioni_SPAD_review,ref:Gulinatti_R_SPAD}. Using a much wider
wavelength separation between signal and pump photons improves the
signal-to-background level by about two orders of magnitude with
respect to Ref.~\onlinecite{ref:Rakher_NPhot_2010}.  Measurements of
photon statistics and two-photon interference before and after
conversion indicate no degradation in purity or wavepacket overlap
of the single photon stream due to the frequency conversion process.
Finally, we show that two spectrally separate transitions of a QD
can be converted to the same wavelength in a single PPLN waveguide,
and present initial measurements demonstrating two-photon
interference of these frequency-converted photons. This represents a
first step towards a resource-efficient approach in which a single
nonlinear crystal acts as a QFC interface that generates
indistinguishable photons from different solid-state
sources~\cite{ref:Sanaka_Yamamoto_PRL}.

The basic experimental system is depicted in Fig.~\ref{fig:Fig1}(a)
and described in detail in the Supplemental
Information~\cite{ref:background_free_QFC_note}. Our single photon
source is an InAs QD in a fiber-coupled, GaAs microdisk optical
cavity~\cite{ref:Srinivasan16} excited by a continuous wave (cw) or
pulsed (50\,MHz repetition rate, 50\,ps pulse width) 780\,nm laser
diode. Spectrally isolated emission from the QD can be studied in
the 980~nm band through photon correlation and two-photon
interference (Hong-Ou-Mandel~\cite{ref:Hong_Ou_Mandel_PRL})
measurements, or else sent to the frequency conversion setup.
Frequency conversion is done by combining a strong, tunable 1550~nm
pump laser with the 980~nm QD signal and coupling them into a PPLN
waveguide. The 600~nm converted signal is spectrally isolated and
sent into either a second photon correlation or Hong-Ou-Mandel
apparatus, to study the photon statistics and two-photon
interference after frequency conversion.

\begin{figure}
\centerline{\includegraphics[width=8.5 cm]{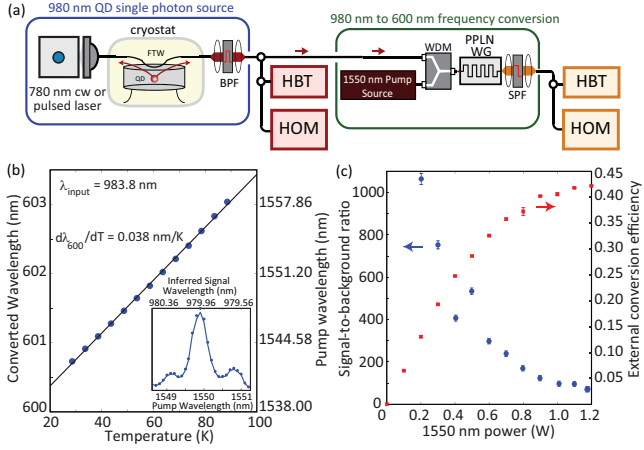}}
\caption{(a) Experimental setup used within this
work~\cite{ref:background_free_QFC_note}. HBT = Hanbury-Brown and
Twiss setup; HOM = Hong-Ou-Mandel interferometer. (b) Converted
600\,nm band wavelength vs. PPLN waveguide temperature. The inset
shows the quasi-phase-matching response of the PPLN waveguide. (c)
Signal-to-background ratio (left y-axis, blue points) and external
conversion efficiency (right y-axis, red points) as a function of
1550\,nm pump power. The external conversion efficiency includes all
losses in the system.} \label{fig:Fig1}
\end{figure}

We characterize the frequency conversion
setup~\cite{ref:background_free_QFC_note} using an attenuated
($\approx$\,30\,fW) 980~nm band laser. First, we measure the
quasi-phase-matching bandwidth of the PPLN waveguide, and find that
it follows the expected sinc$^2$ response~\cite{ref:Fejer_IEEE} with
an inferred bandwidth in the 980~nm band of $\approx$\,0.20\,nm
(inset of Fig.~\ref{fig:Fig1}(b)). Next, we study how the frequency
converted wavelength changes with PPLN waveguide temperature, which
influences phase-matching through thermo-optic and thermal expansion
contributions.  The resulting plot in Fig.~\ref{fig:Fig1}(b)
indicates that the output wavelength can be tuned by
$\approx$\,2\,nm. We have also found that signals between 970\,nm
and $>$\,995\,nm can be converted ($>35~\%$ external conversion
efficiency) by appropriately adjusting the 1550~nm wavelength and
PPLN waveguide temperature. This covers the s-shell emission range
of the QD ensemble, and means that QDs emitting at different
wavelengths (unavoidable due to size/shape/composition dispersion
during growth) can be converted to the same wavelength.

Ideally, QFC should avoid generating noise photons that are
spectrally unresolvable from the frequency-converted quantum state.
Sum- and difference-frequency generation in $\chi^{(2)}$ materials
are background-free in principle~\cite{ref:Kumar}, meaning that
signal photons are directly converted to idler photons without
amplifying vacuum fluctuations. However, other processes, such as
frequency conversion of broadband Raman-scattered pump photons, may
still be a source of noise, as observed in experiments using PPLN
waveguides~\cite{ref:Langrock_Fejer}. To quantify this, the
signal-to-background ratio of the converted signal is measured, and
reveals the fraction of converted photons originating from the
signal rather than noise processes. In previous
work~\cite{ref:Rakher_NPhot_2010}, the signal-to-background was
limited to 7:1, and though use of a pulsed pump removed temporally
distinguishable background noise~\cite{ref:Rakher_PRL_2011}, it did
not improve the signal-to-background level. While better spectral
filtering provides improvement ($>$~10:1 signal-to-background was
reported recently~\cite{ref:Zaske}), it is perhaps more desirable to
suppress the noise source, for example, by increasing the separation
between the signal and red-detuned
pump~\cite{ref:Langrock_Fejer,ref:Pelc,ref:Dong_upconversion}. Here,
our pump-signal separation is nearly 600\,nm, suggesting potentially
significant improvement.

To test this, we measure (Fig.~\ref{fig:Fig1}(c)) the
signal-to-background level by spectrally isolating the 600~nm
conversion band~\cite{ref:background_free_QFC_note} and comparing
the detected counts on the SPAD with and without the presence of the
980\,nm band signal (the SPAD dark count rate of
$\approx$~50~s$^{-1}$ is subtracted to give a detector-independent
metric). We also plot the external conversion efficiency, which
includes all PPLN input/output coupling, free-space transmission,
and spectral filtering losses (detector quantum efficiency is not
included). The signal-to-background level remains above 100 for all
but the highest 1550\,nm pump powers, where the conversion
efficiency has begun to roll off. For the experiments that follow,
we operate with a 35~$\%$ to 40~$\%$ external conversion efficiency
and a signal-to-background level $>$100. As the PPLN incoupling
efficiency is $\approx$~60$~\%$, and the transmission through all
optics after the PPLN waveguide is $\approx$~80$~\%$, the internal
conversion efficiency in the PPLN waveguide is $>$~70$~\%$.

\begin{figure}[t]
\centerline{\includegraphics[width=8.5 cm]{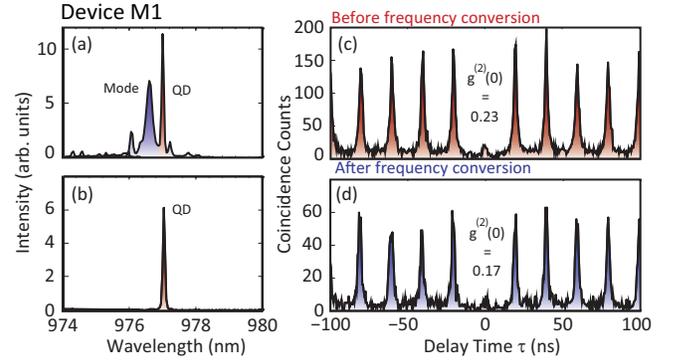}}
\caption{(a) Low-temperature $\mu$-PL spectrum of device M1. Bright
QD emission and cavity mode emission are visible around 977\,nm. (b)
Spectrum of QD emission filtered by a volume Bragg grating. (c)-(d)
Second-order autocorrelation function measurements performed on the
QD emission line before and after frequency conversion.}
\label{fig:Fig2}
\end{figure}

We now present measurements combining frequency conversion with
QD-based single photon sources. We study three devices, M1, M2, and
M3, under pulsed and cw excitation conditions. Pulsed measurements
are a convenient way to judge the temporal distribution of noise
photons produced in the conversion process. Figure~\ref{fig:Fig2}(a)
shows a low temperature (T\,=\,10\,K) micro-photoluminescence
($\mu$-PL) spectrum of device M1 under 780~nm pulsed excitation. A
bright single QD exciton line at 977.04\,nm is visible next to a
cavity mode at 976.65\,nm.  The QD emission line was spectrally
filtered by a volume Bragg grating whose output was coupled to a
single mode fiber (Fig.~\ref{fig:Fig2}(b) shows the filtered QD
emission). Before performing frequency conversion, this filtered
emission was directed to an HBT setup for photon correlation
measurements, the results of which are shown in
Fig.~\ref{fig:Fig2}(c). A strong suppression of the peak at zero
time delay to a value of $g^{(2)}(0)\,=\,0.23\,\pm\,0.04\,<\,0.5$ is
observed. Next, the filtered PL was sent to the frequency conversion
setup, and an auto-correlation measurement was performed on the QD
emission after it was converted to 600\,nm. As shown in
Fig.~\ref{fig:Fig2}(d), the single-photon nature of the QD emission
was preserved during the conversion process, proven by the value of
$g^{(2)}(0)\,=\,0.17\,\pm\,0.03$, and no excess noise from the
frequency conversion process was observed.  In fact, the additional
spectral filtering provided by the quasi-phase-matching process is
the likely cause of the reduction in $g^{(2)}(0)$ after frequency
conversion. 

\begin{figure}[t]
\centerline{\includegraphics[width=8.5 cm]{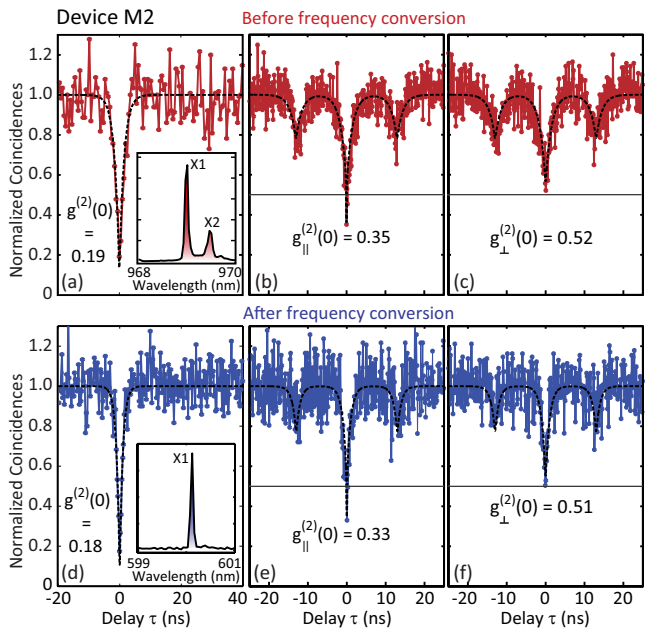}}
\caption{(a) Auto-correlation of the X1 emission line from device M2
under cw excitation ($\mu$-PL spectrum inset). (b) and (c)
Two-photon interference of the X1 line under parallel and orthogonal
polarization configurations of the interferometer arms,
respectively. (d) Auto-correlation of the X1 line after frequency
conversion (frequency converted spectrum inset). (e) and (f)
Two-photon interference of the frequency converted X1 line under
parallel and orthogonal polarization configurations of the
interferometer arms, respectively. The dashed lines are fits to the
experimental data~\cite{ref:background_free_QFC_note}, and the solid
line marks $g^{(2)}(0)\,=\,0.5$ level.} \label{fig:Fig3}
\end{figure}

Similar measurements were performed under cw excitation on device
M2, whose PL spectrum is shown in the inset to
Fig.~\ref{fig:Fig3}(a). Two bright excitonic lines X1 and X2 are
observed on top of a broad cavity mode around 969.5\,nm.
Figures~\ref{fig:Fig3}(a) and (d) show auto-correlation measurements
performed on the filtered X1 line before and after frequency
conversion to 600\,nm, respectively (See inset of Fig.~\ref{fig:Fig3}
(d) for the PL spectrum of converted signal). Antibunching dips in both
figures ($g^{(2)}_{\text{before}}(0)\,=\,0.19\pm0.01$ and
$g^{(2)}_{\text{after}}(0)\,=\,0.18\pm0.02$) again show that the
single photon nature of QD emission is conserved through the
frequency conversion process.

In many cases, both single photon purity and single photon
indistinguishability~\cite{ref:Santori2} are important.  At the
heart of indistinguishability measurements is two-photon
interference~\cite{ref:Hong_Ou_Mandel_PRL}, which we now show is
preserved in our frequency conversion process. Two-photon
interference under cw excitation was performed using a fiber-based
Mach-Zehnder interferometer~\cite{ref:background_free_QFC_note}
similar to
Refs.~\cite{ref:Kiraz_single_molecule_indistinguishable,ref:Patel_PRL08},
where one interferometer arm contains a 12.5~ns delay and a
polarization rotator. Rotating the polarization of photons from this
arm that are incident on the second beamsplitter of the Mach-Zehnder
reveals the effect of interference on the photon correlations. In
the orthogonal polarization configuration, the interferometer arms
are distinguishable and $g^{(2)}_{\perp}(0)\,=\,0.5$ for a pure
single photon source. On the other hand, in the parallel
polarization configuration, one expects interference between the
photons within their coherence time, leading to
$g^{(2)}_{\parallel}(0)\,=\,0$. Figures~\ref{fig:Fig3}(b) and (c)
show the results of experiments on the X1 emission before frequency
conversion. The antibunching values are
$g^{(2)}_{\parallel}(0)\,=\,0.35\pm0.03$ and
$g^{(2)}_{\perp}(0)\,=\,0.52\pm0.04$, yielding the visibility of
two-photon interference as
$\text{V}=(g^{(2)}_{\perp}(0)-g^{(2)}_{\parallel}(0))/g^{(2)}_{\perp}(0)=0.33\pm0.08.$
The deviation from the ideal value of V\,=\,1 stems from the
non-zero value of $g^{(2)}(0)$ (Fig.\ref{fig:Fig3}(a)) and the time
resolution of the photon correlation setup that is on the order of
the coherence time ($\approx\,100$~ps) of the QD
emission~\cite{ref:Patel_PRL08}. The same experiments were performed
on the X1 emission line after frequency conversion, and
Figs.~\ref{fig:Fig3}(e) and (f) show the results for parallel and
orthogonal polarization configurations, respectively
($g^{(2)}_{\parallel}(0)\,=\,0.33\pm0.03$,
$g^{(2)}_{\perp}(0)\,=\,0.51\pm0.05$). Due to the conservation of
the QD coherence time during the frequency conversion process, we
observed a similar two-photon interference visibility
$\text{V}=0.35\pm0.09$ at 600\,nm.

\begin{figure}[!t]
\centerline{\includegraphics[width=8.5 cm]{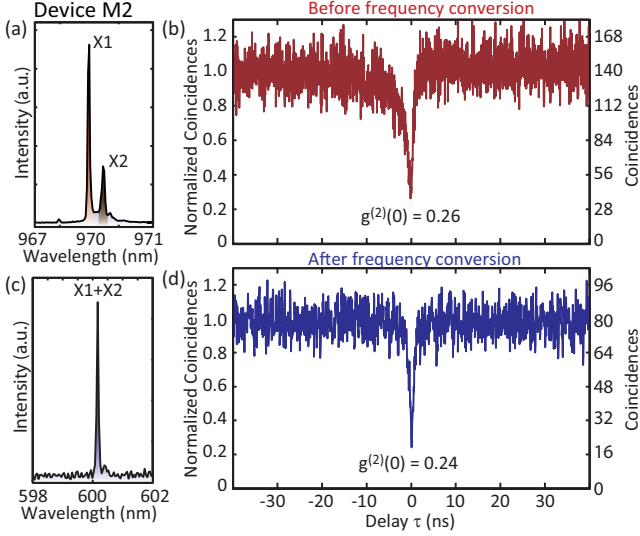}}
\caption{(a) $\mu$-PL spectrum of device M2 under above-band
excitation. (b) Cross-correlation measurement performed on X1 and X2
emission lines. (c) PL spectrum after both lines are converted to
the same wavelength at 600\,nm. (d) Auto-correlation measurement of
the combined frequency converted signal of X1 and X2.}
\label{fig:Fig4}
\end{figure}

As discussed earlier, a wide wavelength range of QD emission within
the 980\,nm band can be efficiently converted to 600\,nm by
controlling the temperature of the PPLN waveguide and the wavelength
of the 1550\,nm pump laser. This enables the conversion of
well-separated emission lines to the same wavelength at 600\,nm. To
demonstrate this, both bright emission lines X1 and X2 from device
M2 ($\mu$-PL spectrum repeated in Fig.~\ref{fig:Fig4}(a)) are
directed to the frequency conversion
setup~\cite{ref:background_free_QFC_note}, together with two
1550\,nm cw pump lasers whose wavelengths are optimized for
efficient conversion of the two 980~nm band signals (which are
separated by $\approx$~0.5~nm). Figure~\ref{fig:Fig4}(c) shows the
PL spectrum of the total converted signal at 600\,nm, where the
converted signals of the individual X1 and X2 lines are spectrally
overlapped (within the spectrometer's resolution
$\approx\,40\,\mu$eV).

To better understand the nature of the measured emission lines, a
cross-correlation measurement was performed before frequency
conversion, where the spectrally filtered X1 and X2 lines were sent
to the stop and start channels of the HBT setup, respectively. As
shown in Fig.~\ref{fig:Fig4}(b), a strong asymmetric antibunching
dip is observed with $g^{(2)}(0)\,=\,0.26\pm0.02$. The antibunching
shows that both emission lines originate from the same QD, while the
asymmetry is related to the radiative dynamics within the QD. The
faster recovery time for $\tau\,>\,0$ can be explained if X1 and X2
arise from neutral and charged excitonic emission,
respectively~\cite{ref:Kiraz3}. This effect arises because emission
of the charged exciton X2 leaves the QD with a single charge, so
that subsequent emission in the neutral exciton state X1 requires
capture of only a single (opposite) charge. This yields a much
faster recovery time than that needed to obtain three charges in the
QD, which sets the recovery time for $\tau<0$.

Next, autocorrelation was performed on the total converted signal at
600 nm, the result of which is shown in Fig.~\ref{fig:Fig4}(d). As
expected, a strong antibunching dip with
$g^{(2)}(0)\,=\,0.24\pm0.02$ is observed. In contrast to the
cross-correlation measurement before conversion, the antibunching
dip now has a symmetric shape. This arises from the fact that both
QD states were converted within a single PPLN waveguide, so that in
the subsequent HBT measurement, the start and stop channels are fed
by the same signal at 600\,nm, which was composed of both X1 and X2
emission lines. This mixing of the signals going into the start and
stop channels removes the asymmetry observed in the
cross-correlation measurement before frequency conversion
(Fig.~\ref{fig:Fig4}(b)).

\begin{figure}
\centerline{\includegraphics[width=8.5 cm]{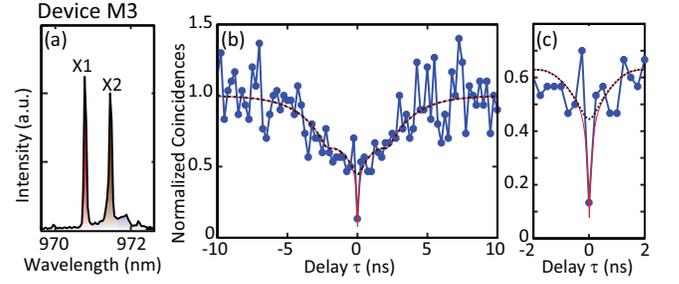}}
\caption{(a) $\mu$-PL spectrum of device M3 under above-band
excitation. Two bright excitonic emission lines (named X1 and X2)
are observed with nearly equal intensity. (b) Two-photon
interference of the combined X1 and X2 signal after both lines are
frequency converted to the same wavelength at 600~nm and measured in
the parallel polarization configuration.  (c) Zoom-in near the
central dip of part (b). The solid red line is a fit to the data,
while the black dashed line corresponds to the orthogonal
polarization configuration.
$g^{(2)}_{\parallel}(0)<g^{(2)}_{\perp}(0)$ is due to the two-photon
interference effect.}\label{fig:Fig5}
\end{figure}

Finally, we consider two-photon interference from two spectrally
distinct QD transitions, as a preliminary step towards using QFC to
generate indistinguishable photons from different QDs, which has
recently been shown through direct tuning of one of the QD
transitions~\cite{ref:Flagg_PRL10,ref:patel_Nphot2010}. We work with
device M3, whose spectrum is shown in Fig.~\ref{fig:Fig5}(a), and
which was chosen because the two excitonic states X1 and X2 have
relatively similar intensities. Cross-correlation
measurements~\cite{ref:background_free_QFC_note} similar to those
described above were performed to confirm that both states come from
the same QD.  After this, the two states were converted to the same
600~nm wavelength as above, and the combined frequency converted
signal was sent into a Mach-Zehnder interferometer similar to that
used earlier.  Data from the parallel polarization configuration is
shown in Fig.~\ref{fig:Fig5}(b)-(c), where the effect of
interference on the photon correlations is seen in the narrow dip at
zero time delay, which reaches a value of
$g^{(2)}_{\parallel}(0)\,=\,0.13\pm0.04$. In comparison, the minimum
calculated value (assuming a pure single photon source and infinite
timing resolution) for the orthogonal (non-interfering) polarization
configuration in our setup~\cite{ref:background_free_QFC_note} is
$g^{(2)}_{\perp}(0)\,=\,0.36$.  This is smaller than the typical
value of 0.5~\cite{ref:Patel_PRL08} due to the delay $\Delta
\tau$=2.2~ns between the interferometer arms, which is comparable to
the average radiative lifetime $T_{1}$=1.7~ns of the two states.
Taking into account the non-zero value $g^{(2)}(0)=0.10$ and the
finite timing resolution of the setup,
$g^{(2)}_{\perp}(0)\,=\,0.45\pm0.04$ is
estimated~\cite{ref:background_free_QFC_note}, far exceeding the
measured value $g^{(2)}_{\parallel}(0)\,=\,0.13\pm0.04$, and
indicating the significant effect of two-photon interference from
the two frequency-converted QD states.

In summary, we have demonstrated background-free quantum frequency
conversion of single photons emitted from a quantum dot.  Photons at
980\,nm are converted to 600\,nm with a signal-to-background larger
than 100 and external conversion efficiency of 40\,\%. We confirm
that single photon purity and wavepacket interference are preserved
during frequency conversion, and demonstrate that spectrally
distinct QD emission lines can be converted to the same wavelength
in the PPLN waveguide.  The ability to use a single frequency conversion
unit to erase spectral distinguishability in solid-state quantum emitters
can be valuable in the development of scalable, chip-based photonic quantum information
devices.

We thank Edward Flagg for information on volume Bragg gratings and
Lijun Ma and Xiao Tang for discussions about PPLN waveguides. S.A.
and I.A. acknowledge support under the Cooperative Research
Agreement between the University of Maryland and NIST-CNST, Award
70NANB10H193.

\onecolumngrid
\newpage
\section{Supplementary Information}

\section{Experimental Setups}

\begin{figure}[b]
\centerline{\includegraphics[width=\linewidth]{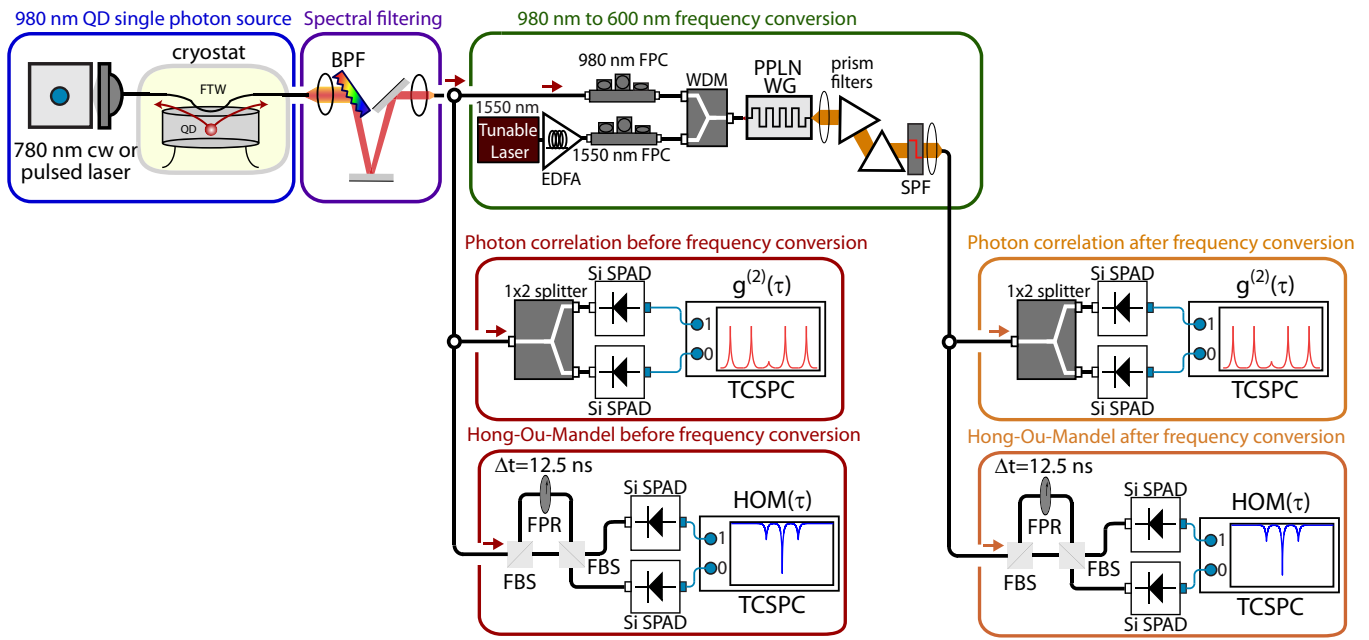}}
\caption{(a) Detailed schematic of the experimental setups used
within this work to demonstrate preservation of photon statistics
and two-photon interference during frequency conversion from the
980~nm band to 600~nm. QD = quantum dot; FTW = fiber taper
waveguide; BPF = bandpass filter; EDFA = erbium-doped fiber
amplifier; FPC = fiber polarization controller; WDM = wavelength
division multiplexer; SPF = short-pass filter; PPLN WG =
periodically-poled lithium niobate waveguide; SPAD = single photon
avalanche diode; FBS = fiber-coupled beamsplitter; FPR =
fiber-coupled polarization rotator; TCSPC = time-correlated single
photon counting board.} \label{fig:SFig1}
\end{figure}

Figure~\ref{fig:SFig1} shows a detailed schematic of the
experimental setup used for photon correlation and two-photon
interference measurements before and after frequency conversion of a
single QD state. The individual sub-systems within this setup are
described below.

\subsection{Quantum dot single photon source}
Our source of single photons consists of an InAs quantum dot (QD) in
a 190~nm thick, 2.9~$\mu$m diameter GaAs microdisk cavity. In this
work, the coupling between the modes of the cavity and the QD
emission lines is weak, and any radiative rate enhancement or
inhibition is relatively small ($<2$~X). QD emission is out-coupled
from the microdisk cavity using an optical fiber taper waveguide
(FTW), which provides a convenient single mode fiber interface for
subsequent experiments. The sample is cooled in a liquid He flow
cryostat, and a sample temperature between 7~K and 10~K is
maintained during all experiments.

\subsection{Spectral filtering of the QD emission} Emission that is
out-coupled by the FTW is sent into a filtering setup consisting of
a $\approx$\,0.2~nm bandwidth volume reflective Bragg grating (BPF
in Fig.~\ref{fig:SFig1}) whose input is coupled to single mode
optical fiber and output is coupled to polarization maintaining (PM)
single mode fiber. Quarter- and half-wave plates and a polarizing
beamsplitter are placed prior to the PM fiber input, to ensure that
light is linearly polarized along the slow-axis of the fiber. The
typical throughput of the filtering setup is between 50$~\%$ and
60$~\%$.

\subsection{Frequency conversion}
In the frequency conversion setup, we use a wavelength division
multiplexer (WDM) to combine the spectrally-filtered QD emission
with a strong (few hundred mW) 1550\,nm pump signal that is
generated by an external cavity tunable diode laser and erbium-doped
fiber amplifier (EDFA). Fiber polarization controllers (FPCs) are
used to adjust the polarization state of both the 980\,nm and
1550\,nm beams. The combined signal and pump are coupled into a
2\,cm long, 5~$\%$ MgO-doped PPLN waveguide whose temperature can be
adjusted between 25~$^{\circ}$C and 90~$^{\circ}$C with
0.1~$^{\circ}$C resolution. Light is coupled into the waveguide
through a cleaved single mode optical fiber that is controlled by a
3-axis open-loop piezo stage and has a mode-field diameter of
5.8\,$\mu$m at 980\,nm. The coupling is optimized for the 980\,nm
band signal, at the expense of the 1550\,nm pump (additional pump
power compensates for the 1550\,nm coupling inefficiency). Light
exiting the PPLN waveguide is collimated and sent through two
dispersive prisms and two 750\,nm short pass edge filters (SPFs) to
eliminate residual 1550\,nm pump photons and frequency doubled
775\,nm pump photons from the signal.  After filtering, the upconverted
signal is usually coupled into a PM single mode fiber. We place a polarizing
beamsplitter prior to the PM fiber output to ensure that light is linearly
polarized along the slow-axis of the fiber (waveplates are not needed because
the output of the PPLN waveguide is very close to linearly polarized).

The external conversion efficiency of the setup is as high as
$\approx$40~$\%$, including all losses in the system other than the
final single mode fiber coupling, which has a typical efficiency of
30~$\%$. In some experiments (i.e., ones in which the spatial mode
profile of the light is unimportant), multimode fiber coupling is
preferred, due to its higher coupling efficiency of 85~$\%$. The
overall detection efficiency of the frequency conversion system is
given as the product of the external conversion efficiency, the
fiber coupling efficiency, and detector quantum efficiency
($\approx$\,67~$\%$ at 600~nm for a thick Si single-photon avalanche
diode).

\subsection{Si single-photon avalanche diode (SPADs)}

Three different types of Si single-photon avalanche diodes (SPADs)
are used in the experiments, depending on the requirements on
detection efficiency and timing resolution.  Thick Si SPADs used in
this work have a detection efficiency of $\approx$12.5~$\%$
(67~$\%$) at 980~nm (600~nm) and a timing jitter $>$500~ps, and are
used in experiments in which high timing resolution are not needed.
Thin Si SPADs~\cite{ref:Ghioni_SPAD_review} have a detection
efficiency of $\approx$2~$\%$ (45~$\%$) at 980~nm (600~nm), and a
timing jitter of $\approx$50~ps.  Newly-developed red-enhanced Si
SPADs~\cite{ref:Gulinatti_R_SPAD} have a detection efficiency of
$\approx$6~$\%$ (55~$\%$) at 980~nm (600~nm), and a timing jitter of
$\approx$100~ps.

As evident from these efficiency and timing jitter values, an
advantage of using frequency conversion to the 600~nm band is the
potential to simultaneously achieve high quantum efficiency
($>40~\%$) and low timing jitter ($\approx~50$~ps), which is
currently not possible in the 980~nm band using Si SPAD technology.

\subsection{Pulsed photon correlation measurement}
In pulsed measurements, the QD is pumped with a gain-switched 780~nm
laser diode operating with a 50\,MHz repetition rate and 50\,ps pulse width.
Prior to frequency conversion, spectrally-isolated QD emission
is split using a 1x2 fiber splitter, and each output of the splitter is
sent to a thick Si SPAD.  The SPAD outputs are sent to a time-correlated
single photon counting (TCSPC) system with a bin size of 512~ps, and histogram
data is acquired to measure the second-order intensity correlation function $g^{(2)}(\tau)$.

Measurements after frequency conversion are performed in a similar fashion, with the output
of the frequency conversion setup directed to a 1x2 fiber splitter, and subsequently, into
thick Si SPADs and the TSCPC.

\subsection{Continuous wave photon correlation measurement}
Continuous wave (cw) measurements are performed in a manner similar to that
used for pulsed measurements, with the following exceptions: (1) a 780~cw
laser is used to excite the QD. (2) High timing resolution SPADs and a smaller
TCSPC histogramming bin width are used.  For measurements before frequency
conversion, two red-enhanced Si SPADs are used, and the bin size of the
TCSPC is set to 256~ps.

For measurements after frequency conversion, one red-enhanced Si SPAD and one
thin Si SPAD are used, and the bin size of the TCSPC is set to 256~ps.

\subsection{Single state two-photon interference measurement}

Two-photon interference measurements are performed using
Mach-Zehnder interferometers for the 980~nm band (before conversion)
and 600~nm band (after conversion).  Each interferometer starts with
a PM-fiber-coupled beamsplitter (FBS in Fig.~\ref{fig:SFig1}). One
output of the FBS is connected directly into the first input port of
a second FBS, while the other output goes through a fiber-coupled
polarization rotator (FPR) before being connected to the second
input port of the second FBS. The FPR can be rotated to allow both
maximum interference (parallel configuration) or minimum
interference (orthogonal configuration), and its input and output
cables provide a delay of 12.5~ns with respect to the other arm of
the interferometer.

For measurements done before frequency conversion, the outputs of
the 980~nm band Hong-Ou-Mandel setup are detected with the two
red-enhanced Si SPADs.  For measurements after frequency conversion,
one red-enhanced Si SPAD and one thin Si SPAD are used, and the
outputs are sent to the TCSPC. Data is acquired in a time-tagged,
time-resolved mode in which the photon arrival times from each
channel are recorded with 4~ps timing resolution and the bin size
set during subsequent data analysis. Here, a bin size of 125~ps was
used for measurements before and after conversion.

\subsection{Photon correlation measurements using two state frequency conversion}

\begin{figure}
\centerline{\includegraphics[width=0.8\linewidth]{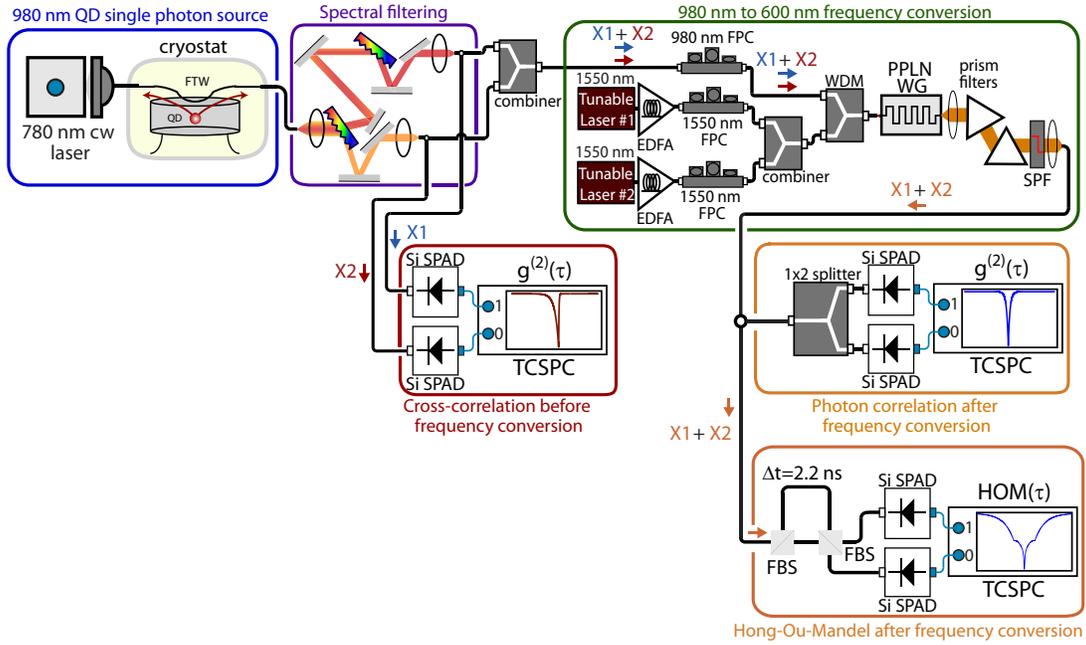}}
\caption{Detailed schematic of the experimental setup used for
frequency conversion of two spectrally-separated QD states to the
same wavelength.} \label{fig:SFig2}
\end{figure}

The experimental setup used for studying the frequency conversion of
two spectrally-separate QD states is shown in Fig.~\ref{fig:SFig2}.
Prior to conversion, we perform cross-correlation measurements by
spectrally isolating lines X1 and X2 using reflective volume
gratings whose outputs are coupled to single mode optical fibers.
The fibers are connected to red-enhanced Si SPADs, whose outputs are
fed into the TCSPC.  Data is acquired in time-tagged, time-resolved
mode and a bin size of 50~ps was used in analyzing the data.

To frequency convert states X1 and X2 to the same wavelength in the
600~nm band, two 1550~nm band external cavity diode lasers are
separately amplified and combined with lines X1 and X2 in the PPLN
waveguide.  Each 1550~nm laser is tuned to maximize the frequency
conversion of the corresponding QD state, and the output of the PPLN
waveguide is spectrally resolved to determine the precise wavelength
to which light is converted.  Fine tuning of the 1550~nm wavelengths
is performed to ensure that both states are converted to the same
wavelength (within the resolution of the grating spectrometer). This
adjustment can come at the expense of conversion efficiency (in
practice, we sacrifice on the conversion efficiency of state X1,
since it is brighter).

The combined frequency converted light from states X1 and X2 is
coupled into a single mode PM fiber and split using a 1x2 fiber
splitter, with the arms of the splitter hooked up to a red-enhanced
Si SPAD and a thin Si SPAD. Data is acquired in time-tagged,
time-resolved mode and a bin size of 50~ps was used in the data
analysis.

\subsection{Two-photon interference measurement using two state frequency conversion}

Two-photon interference of two QD states follows the same procedure
as above, where both states are frequency converted to the same
wavelength in a single PPLN waveguide.  The combined frequency
converted light from the two states is sent to a Mach-Zehnder
interferometer (Fig.~\ref{fig:SFig2}), and the outputs of the
interferometer are hooked up to a red-enhanced Si SPAD and a thin Si
SPAD.  Data is acquired in time-tagged, time-resolved mode, and due
to the relatively low photon count rates in this measurement (more
details are given in the final section below), a bin size of
250~ps was used in analysis of the data.

\section{Measurements and Data Analysis}

\subsection{Quasi-phase-matching response and temperature tuning}

We measure the quasi-phase-matching bandwidth of the PPLN waveguide
as follows. The weak signal produced by a narrow linewidth
($<$5\,MHz) laser is fixed at 983.8~nm, and the temperature of the
PPLN waveguide is set to 58.8~$^{\circ}$C. Next, the 1550\,nm pump
power is set to $\approx$\,800\,mW, close to the value at which we
achieve optimal conversion efficiency. We then scan the 1550\,nm
pump wavelength while monitoring the frequency converted 600\,nm
band signal on a thick Si SPAD.  This generates the plot shown in
the inset to Fig.~1(b).  We infer the 980~nm quasi-phase-matching
bandwidth based upon energy conservation and the 1550~nm
wavelengths.

Adjusting the temperature of the PPLN waveguide provides a mechanism
to tune the output wavelength of the frequency conversion process.
At each PPLN waveguide temperature, we keep the input signal fixed
at 983.8\,nm and scan the 1550~nm laser, recording the 1550~nm
wavelength at which the conversion efficiency is maximized. Using
this 1550~nm wavelength and the requirement of energy conservation
produces the plot of frequency converted wavelength against
temperature in Fig.~1(b).

\subsection{Signal-to-background measurements}

Signal-to-background measurements are performed using a thick Si
SPAD placed after the prisms and SPFs, and with a 1550~nm band pump
beam whose average power at the input of the 980/1550~nm WDM varies
between 0~W and 1.2~W.  The converted signal was measured using a
weak 980~nm band input signal produced by an attenuated laser.  The
output average power level is set to be consistent with that of the
QD single photon sources under investigation ($\approx$30~fW), while
the background level was determined by switching off the 980~nm
laser.  The dark count rate of the SPAD ($\approx$50~s$^{-1}$) was
subtracted from both the converted signal and background levels, to
provide a metric that is independent of the characteristics of the
detector. The error bars in Fig. 1(c) are one standard deviation
values and are due to fluctuations in the detected count rate on the
SPAD.

\subsection{Lifetime measurements}

Lifetime measurements reveal a radiative lifetime of
$T_{1}\approx$~1~ns for device M1, while devices M2 and M3 show
longer lifetimes of $\approx$1.5~ns and $1.7$~ns, respectively.
These measurements are performed using the 780~nm pulsed laser, with
the electronic trigger output of the laser fed to the first channel
of the TCSPC, and the detected signal from a red-enhanced Si SPAD
fed to the second channel.

\subsection{$g^{(2)}(\tau)$ measurements}

Measurement of the second-order correlation function $g^{(2)}(\tau)$
are performed using the setups described above. In pulsed
measurements, the $g^{(2)}(0)$ value is determined by comparing the
integrated area of the peak around time zero to the average area of
the peaks away from time zero. The uncertainty on this value is
given by the standard deviation in the area of the peaks away from
time zero.  $g^{(2)}(\tau)$ under cw excitation is normalized to the
coincidence rate at long time delays ($\tau>500$~ns).  The
uncertainty in $g^{(2)}(0)$ is due to the fluctuation in coincidence
rate for $\tau>500$~ns, and is the one standard deviation value.
Measurements are run for a long enough time for the Poissonian
levels to reach a minimum of 80 coincidences.

\subsection{Single state two-photon interference measurements}

Two-photon interference experiments on device M2 are performed using
the setups described above. The correlation functions at the output
of the Mach-Zehnder interferometers under parallel and orthogonal
polarizations of the photons, $g^{(2)}_{\parallel}(\tau)$ and
$g^{(2)}_{\perp}(\tau)$, are normalized to the coincidence rate at
long time delays ($\tau>500$~ns).  The uncertainties in
$g^{(2)}_{\parallel,\perp}(0)$ are due to the fluctuations in the
coincidence rate for $\tau>500$~ns, and are listed as the one
standard deviation value.  Measurements are run for a long enough
time for the Poissonian levels to reach a minimum of 100
coincidences.

To aid in the analysis of the two-photon interference data, we
follow the procedure of Refs.~\onlinecite{ref:Patel_PRL08} and
~\onlinecite{ref:Kiraz_single_molecule_indistinguishable}. First, we
consider that in the limit of no timing resolution limitations, the
second-order correlation function $g^{(2)}(\tau)$ is of the form:

\begin{equation}
g^{(2)}(\tau) = 1-\alpha e^{(-|\tau|/\tau_{r})},
\label{eq:eq1}
\end{equation}

\noindent where $\tau_{r}$ is the radiative lifetime of the QD
state, and $\alpha$ accounts for potential multi-photon
contributions to the data (i.e., that lead to a non-zero value of
$g^{(2)}(0)$).  The timing resolution of the SPADs and TCSPC are
taken into account by convolving eq.~(\ref{eq:eq1}) with the SPAD
instrument response function and binning the result over a time
window equal to the time bin set on the TCSPC.

Using the above approach, we fit the cw $g^{(2)}(\tau)$ data, as
shown in the dashed lines in Fig. 3(a) and (d).  The values we
extract for $\alpha$ ($\approx$0.1) and $\tau_{r}$ ($\approx$1.5~ns)
are then used with eq.~(\ref{eq:eq1}) in formulas for
$g^{(2)}_{\parallel}(\tau)$ and
$g^{(2)}_{\perp}(\tau)$~[\onlinecite{ref:Patel_PRL08}]:

\begin{equation}
g^{(2)}_{\parallel}(\tau) = 4(T_{1}^2+R_{1}^2)R_{2}T_{2}g^{(2)}(\tau)+ \\
4R_{1}T_{1}[T_{2}^2g^{(2)}(\tau-\Delta\tau)+R_{2}^2g^{(2)}(\tau+\Delta\tau)](1-ve^{-2|\tau|/\tau_{c}})),
\label{eq:eq2}
\end{equation}

\begin{equation}
g^{(2)}_{\perp}(\tau) = 4(T_{1}^2+R_{1}^2)R_{2}T_{2}g^{(2)}(\tau)+ \\
4R_{1}T_{1}[T_{2}^2g^{(2)}(\tau-\Delta\tau)+R_{2}^2g^{(2)}(\tau+\Delta\tau)],
\label{eq:eq2}
\end{equation}

\noindent $R$ and $T$ are the reflection and transmission intensity
coefficients at the two beamsplitters, which we assume to be 0.5 for
all of the beamsplitters in use.  $v$ is the spatial overlap of the
photon wavepackets at the second beamsplitter, which we assume to be
unity, and $\Delta \tau$ is the delay in the fiber-coupled
Mach-Zehnder interferometer, which is $\approx$12.5~ns. Accounting
for the timing resolution of the SPADs and TCSPC in the same way as
described above for $g^{(2)}(\tau)$, we produce the dashed line
curves shown in Fig. 3(b),(c) and (e),(f).  The extracted coherence
time $\tau_{c}=100$~ps is consistent with independent measurements
of coherence time for QDs under 780~nm excitation on this sample,
which show a range of values between $\approx$75~ps and 250~ps.

Finally, the visibility of the two-photon interference is given by:

\begin{equation}
\text{V}=(g^{(2)}_{\perp}(0)-g^{(2)}_{\parallel}(0))/g^{(2)}_{\perp}(0).
\label{eq:eq3}
\end{equation}

\subsection{Two state two-photon interference measurements}
\label{subsec:two_state_two_photon}

\begin{figure}
\centerline{\includegraphics[width=0.5\linewidth]{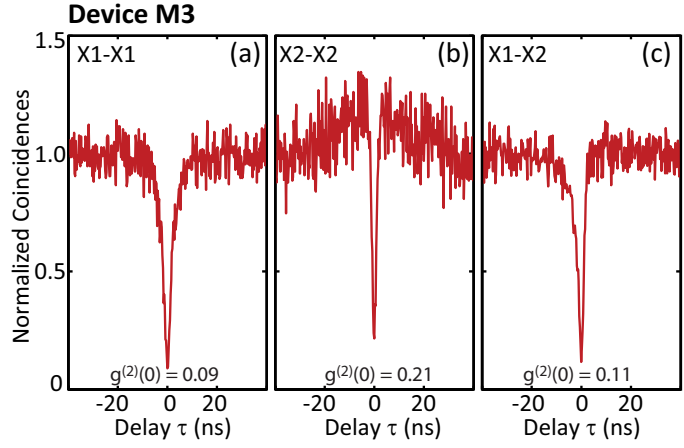}}
\caption{Data for device M3 before frequency conversion. (a)
Autocorrelation of state X1, (b) Autocorrelation of state X2, (c)
Cross-correlation between states X1 and X2.} \label{fig:SFig3}
\end{figure}

Autocorrelation and cross-correlation data from states X1 and X2 of
device M3 are shown in Fig.~\ref{fig:SFig3}, and establish that both
states come from the same QD.  For the X1 and X2 autocorrelation,
$g^{(2)}(0)=0.10\pm0.02$ and $g^{(2)}(0)=0.21\pm0.02$, respectively,
while the X1-X2 cross-correlation has $g^{(2)}(0)=0.11\pm0.02$.  The
difference in $g^{(2)}(0)$ values between X1 and X2 arises from the
relative position of the two states with respect to a broad cavity
mode, which is more closely aligned with state X2.  This also
influences the radiative lifetime of the two states, as clearly
visible in the correlation data in Fig.~\ref{fig:SFig3}.

Two-photon interference of the frequency converted signal from the
states X1 and X2 is performed using the 600~nm Mach-Zehnder
interferometer (Fig.~\ref{fig:SFig2}). To limit losses, the
fiber-coupled polarization rotator is removed in the parallel
polarization configuration (since the polarization is correctly
aligned without it), and the resulting delay between the two arms of
the interferometer is 2.2~ns.  In addition, a bin size of 250~ps is
used in analyzing the time-tagged, time-resolved data, which was
acquired for 2.5~h to produce the data in Fig.~5(b)-(c).

The data is fit using the same approach as described in the single
state two-photon interference section, but now with $\Delta
\tau$=2.2~ns.  Because $g^{(2)}(\tau=2.2~\text{ns})<1$ (in part due
to the relatively long lifetimes of the states involved, with
$T_{1}\gtrsim1.5$~ns), the value of $g^{(2)}(0)$ in the
perpendicular polarization configuration, $g^{(2)}_{\perp}(0)$, can
be less than 0.5, which is what one typically predicts for this
non-interfering configuration in the case of a pure single photon
source~\cite{ref:Patel_PRL08}.  Non-zero multiphoton probability
(i.e., $g^{(2)}(0)>0$) will also influence $g^{(2)}_{\perp}(0)$. Our
best estimates of $T_{1}=1.7$~ns, $g^{(2)}(0)=0.1$, and
$\tau_{c}$=200~ps yield the red solid curve (parallel polarization
configuration) and black dashed line (orthogonal polarization
configuration) in Fig.~5(b).  This gives
$g^{(2)}_{\perp}(0)\,=\,0.45\pm0.04$, and even if infinite timing
resolution of the detectors and $g^{(2)}(0)=0$ are assumed,
$g^{(2)}_{\perp}(0)\,=\,0.36$ . The significant reduction measured
in the parallel polarization configuration, with
$g^{(2)}_{\parallel}(0)\,=\,0.13\pm0.04$, is thus attributed to the
two-photon interference effect.

Future improvements to the setup will likely require improved
optical losses. Extra losses within the setup relative to the
two-photon interference of a single QD state include: (1) The
necessity to achieve precise spectral matching of the upconverted
wavelength, which requires fine tuning of the two 1550~nm pump
lasers away from the peak conversion wavelengths, (2) The limited
1550~nm band pump power available due to the directional coupler
used to combine the two pump signals, which limits the power in any
one pump field to $\approx$~500~mW and prevents operation at the
peak of the external conversion efficiency curve of Fig.1(c), and
(3) The directional coupler used to combine the X1 and X2 emission
after spectral filtering, which automatically adds 3~dB of loss due
to the unused output port.  Improvement in the losses of the
fiber-coupled beamsplitters ($>$3~dB total) and the single mode
fiber coupling at the output of the PPLN waveguide (5.2~dB) (also
present in the single state two-photon interference experiment)
would also help improve the signal-to-noise level in the
measurements.

\end{document}